\documentclass[openacc]{rstransa}

\usepackage[latin1]{inputenc}

\begin{document}

\title{Out-of-Equilibrium Spin Transport in Mesoscopic Superconductors}

\author{C.H.L. Quay $^{1}$ and M. Aprili$^{1}$}

\address{$^{1}$Laboratoire de Physique des Solides, CNRS, Univ. Paris-Sud, University Paris-Saclay, 91405 Orsay Cedex, France\\}

\subject{low-temperature physics, materials science, mesoscopics, nanotechnology, quantum physics, spintronics}

\keywords{superconductivity, out-of-equilibrium, spintronics, mesoscopic physics}

\corres{M. Aprili\\
\email{marco.aprili@u-psud.fr}}

\begin{abstract}

The excitations in conventional superconductors, Bogoliubov quasiparticles, are spin-1/2 fermions but their charge is energy-dependent and, in fact, zero at the gap edge. Therefore, in superconductors (unlike normal metals) the spin and charge degrees of freedom may be separated. In this article, we review spin injection into conventional superconductors and focus on recent experiments on mesoscopic superconductors. We show how quasiparticle spin transport and out-of-equilibrium spin dependent superconductivity can be triggered using the Zeeman splitting of the quasiparticle density of states in thin-film superconductors with small spin-mixing scattering. Finally, we address the spin dynamics and the feedback of quasiparticle spin imbalances on the strength of the superconducting energy gap.

\end{abstract}

\begin{fmtext}

\end{fmtext}

\maketitle

\section{Introduction}

Out-of-equilibrium superconductivity has received a lot of attention since the 1970s~\cite{1,2}, with the Bardeen-Cooper-Schrieffer (BCS) theory of superconductivity providing a framework for describing the excitations of a `conventional' superconducting condensate of singlet Cooper pairs~\cite{3}. The advent of micro- then nanotechnology and their continuous improvement greatly facilitated the fabrication of devices and circuits with dimensions on the order of the excitations' relaxation lengths of 1--100$\mu$m~\cite{4}. Out-of-equilibrium states have thus become achievable with small external perturbations, e.g. external voltage biases, and also more easily detected. In parallel, spin-dependent superconducting proximity effects in hybrid devices at equilibrium were discovered; this was enabled by progress in engineering superconductor/(ferro)magnet interfaces. Oscillations of the superconducting order parameter, ferromagnetic $\pi$-junctions~\cite{4_bis} as well as long-range odd-frequency triplet correlations~\cite{23} have been observed. These results belied received wisdom about the incompatibility between superconductivity and magnetism, and called for a better understanding of spin transport and polarization in conventional superconductors.

Work on out-of-equilibrium superconductivity has concentrated mainly on the energy and charge excitation modes associated, respectively, with heat and charge transport~\cite{5} and their applications in cryogenic detection of radiation and thermometry~\cite{6}. These excitations correspond to an excess of quasiparticles with respect to thermal ones (energy mode)~\cite{parker} and an imbalance between the hole-like and electron-like excitations (charge mode)~\cite{7,owen}. Both energy and charge modes are spin degenerate, though quasiparticles are spin-1/2 fermions and obey Fermi-Dirac statistics at equilibrium --- their distribution function is then given by  $ f(E)= 1/\big[1+\exp(E/k_BT)\big]$, with $E$ the energy, $T$ the temperature and $k_B$ Boltzmann's constant. (The Fermi energy $E_F$ is taken to be equal to zero.)

The energy mode can be excited by charge-neutral perturbations such as electromagnetic radiation whose frequency is larger than the superconducting energy gap $\Delta$; the absorption of such radiation breaks pairs and creates quasiparticules~\cite{8}. The charge mode, on the other hand, can be excited by injecting charged carriers (i.e. electrons or holes) from a normal electrode into a superconductor, where they become quasiparticles~\cite{9}. As quasiparticles are not instantaneously converted into Cooper pairs, their chemical potential is shifted up or down with respect to that of Cooper pairs. This has been measured as a voltage drop between the superconductor and a normal electrode in contact with the superconductor via a tunnel barrier~\cite{10,11}. If electrons or holes are injected at energy $\ |E|>\Delta$, both charge and energy modes are excited. The relaxation time for the energy mode is the inelastic scattering time~\cite{6} while charge imbalance relaxes over the pair-breaking time (or equivalently quasiparticle recombination time)~\cite{12}.

In a visionary theoretical paper published in 1976, A. Aronov~\cite{13} introduced the concept of spin injection into superconductors. The main idea of his paper was to use spin injection to produce an internal magnetic field in order to perform an NMR experiment in the superconducting state. (The Meissner effect prevents external magnetic fields from penetrating into the volume of superconductors.)

In a pioneering experiment, Johnson et al.~\cite{14} showed in 1985 that spin injection from a ferromagnetic electrode into a normal metal is possible by applying a voltage bias across the interface between the two. The out-of-equilibrium  magnetization created in the normal metal is detected electrically, by measuring the voltage between it and a second ferromagnetic electrode~\cite{15}. This `nonlocal' signal is directly proportional to the shift in the chemical potential, $\mu_s$, of spin up (down) electrons due to spin accumulation~\cite{16}, \cite{17}. (spin up and down chemical potentials shift by the same amount, but in opposite directions.) The spin relaxation length measured in high purity light metals (which have low spin-orbit coupling) can reach $100\mu m$~\cite{14} while the spin relaxation time is $\sim$50ns. Subsequently Johnson et al. extended spin injection into a superconductor close to its critical temperature $T_c$~\cite{18}.  This not only raised the question of spin transport in superconductors, the spins being carried by the out-of-equilibrium quasiparticles, but also motivated a search for superconducting states in which singlet Cooper pairs coexist with spin-polarized quasiparticles~\cite{19}.

Among excitation techniques besides those mentioned above, we note in particular charge and spin currents generated by the magnetic losses of the precessing magnetisation of a ferromagnet also called `spin pumping'~\cite{20,21,22,moriyama,xiao,tserkovnyak-rmp,tserkovnyak-prb-08}. In the case of a ferromagnetic insulator in contact with a superconductor, a pure spin current with no charge should be injected~\cite{kajiwara}. In addition, as this `spin pumping' technique injects spins at low energies, in contrast to spin-polarised current injection, it should in principle also result in little or no excitation of the charge mode. Nevertheless, while spin pumping into metals has been demonstrated experimentally, this technique has not been applied to superconductors.

\section{Charge-spin separation and relaxation}

Spin injection from a ferromagnetic electrode as in Johnson's work~\cite{14} is always associated with charge transport and charge imbalance as it arises from a charge current of spin-polarised carriers. However, unlike normal metals, in which charge imbalance cannot exist, in superconductors both charge and spin imbalances are possible in the quasiparticle population. Fundamentally, this is because the superconducting condensate acts as a `charge reservoir', such that quasiparticle charge need not be conserved~\cite{39}. (The total charge of both quasiparticles and Cooper pairs is conserved. Charge imbalance in the quasiparticle population thus means that the quasiparticle chemical potential is shifted with respect to that of the pairs~\cite{7,9,10, 39}). Note, in addition, that the charge of Bogouliobov spin-1/2 quasiparticles is energy dependent: $q(E)=e\sqrt{E^2-\Delta^2}/E$; therefore, while quasiparticles at the gap edge are neutral, high energy quasiparticles are $+e$ (hole-like) or $-e$ (electron-like) excitations~\cite{24}. This results in an effective spin-charge separation in the quasiparticle population as pointed out by Kivelson et al.~\cite{25}. As spin and charge can relax via different mechanisms and thus over different timescales, this leads to a peculiar situation in diffusive out-of-equilibrium superconductors in which charge and spin excitation modes are spatially separated.

A superconductor in good metallic (i.e. high transparency) contact with a ferromagnet experiences a strong suppression of superconductivity in a region up to a few superconducting coherence lengths $\xi$ from the interface, even at equilibrium~\cite{26,28}. (The superconducting coherence length is $\xi = \hbar v_F/ \Delta $ in the clean limit and $\xi= \sqrt{\hbar D/\Delta}$ in the diffusive limit, with $D$ the diffusion constant,  $v_F$ the Fermi velocity and $\hbar$ Planck's constant.) This phenomenon, also known as the inverse proximity effect, is due to the strong pair-breaking effect of the exchange field on singlet Cooper pairs~\cite{27} that leak from the superconductor into the ferromagnet, which feeds back through the boundary conditions on superconductivity at the interface~\cite{28}.

Placing a tunnel barrier between the superconductor and the ferromagnet decouples the spin and Cooper pair reservoirs and suppresses undesirable proximity and inverse proximity effects~\cite{29}, including the appearance of Andreev bound states in the superconducting gap~\cite{blonder}. This also affords a technical advantage: as a tunnel barrier results in a finite interface resistance, the interface can now be voltage biased and thus the quasiparticle injection energy chosen. Injection at energies lower than the superconducting energy gap is not possible due to the lack of quasiparticle states; the bias voltage $V$ must be thus larger than $\Delta$. Finally, tunneling through an insulator preserves spin orientation; this turns out to be essential for efficient spin injection~\cite{30}. In the following, `spin injection'  should be taken to mean spin injection through a tunnel barrier. As the spin and charge degrees of freedom are not directly coupled, we focus mainly on spin imbalance. Charge imbalance phenomena are treated in Ref.s~\cite{31}.

Assuming negligible spin accumulation in the ferromagnetic electrode as well as a uniform temperature for the whole system and no charge imbalance (in the superconductor), the spin, `electrical charge' and `quasiparticle charge' currents across a ferromagnet-insulator-superconductor junction obtained from the Fermi Golden rule read~\cite{32,33}:

\begin{align}\label{currents}
\begin{split}
I_{e}(V) = \frac{2 \pi e}{\hbar} N_0 |\mathcal{T}|^2 \int_{-\infty}^{+\infty} N_{BCS}(E) \Big\{N_{F} \big[f(E-eV)-f(E+eV)\big]-\delta N_{F} \big[f(E-\mu_s)-f(E+\mu_s)\big]\Big\}dE,\\
I_{s}(V) = \pi N_0 |\mathcal{T}|^2 \int_{-\infty}^{+\infty} N_{BCS}(E) \Big\{\delta N_{F} \big[f(E-eV)-f(E+eV)\big]-N_{F} \big[f(E-\mu_s)-f(E+\mu_s)\big]\Big\}dE, \\
I_{QP}(V) = \frac{\pi }{\hbar} N_0 |\mathcal{T}|^2 \int_{-\infty}^{+\infty}q^2(E) N_{BCS}(E) \Big\{N_{F} \big[f(E-eV)-f(E+eV)\big]+ \delta N_{F} \big[f(E+\mu_s)-f(E-\mu_s)\big]\Big\}dE.
\end{split}
\end{align}

Here $V$ is the voltage (applied or measured) across the junction; $e$ the electron charge, $N_0$ the density of states of the superconductor at the Fermi level when it is in the normal state; $N_{BCS}(E)= \lvert E\rvert/\sqrt{(E^2-\Delta^2)}$ the BCS quasiparticle density of states; $\mu_s$ a chemical potential shift (of opposite sign for opposite spins ); $q(E)$ the quasiparticle charge; and $|\mathcal{T}|^2$ the junction transmission assumed to be identical for each spin. $N_F = N_{F \uparrow} + N_{F \downarrow}$ and $\delta N_F = N_{F\uparrow} - N_{F\downarrow}$ account for the different density of states, $N_{F\uparrow}$ and  $N_{F\downarrow}$ at the Fermi level for spin up and down in the ferromagnet . (The polarisation of the ferromagnet is encoded in $\delta N_F$; thus, in the case where the ferromagnet is replaced by a normal metal electrode $\delta N_F = 0$.) The assumption of a uniform system temperature implies fast thermalization~\cite{35}.

The difference between the electrical charge current and the quasiparticle charge current is carried by the superconducting condensate~\cite{34}. These equations apply equally to both injector (`local') and detector (`nonlocal') junctions: At the injector the currents are determined by the applied voltage $V_L$, while at the latter a voltage $V_{NL}$ is measured in presence of spin imbalance. $V_{NL}$ can be found by inverting the  Equations~\ref{currents} and imposing the condition $I_{e}(V_{NL})=0$, i.e. zero charge current through the detector as is usual for a voltage probe.

Spin accumulation appears in the superconductor --- in a dynamic equilibrium between injection and relaxation --- and can be described by: $dS(t)/dt= I_s(t)-S(t)/\tau_s$, with $S(t)$ the spin imbalance and $\tau_s$ the spin imbalance relaxation time. In principle, this equation is coupled to Equations~\ref{currents} as $\mu_s$ is directly linked to $S(t)$, and the whole system of equations must be solved self-consistently. However if the superconductor is only slightly out-of-equilibrium (i.e. $\mu_s \ll V_L$, $\Delta$) $\mu_s$ can be neglected in Equations~\ref{currents} and in the stationary limit (i.e. $dS/dt=0$) we obtain $S=I_s \tau_s=P I_e\tau_s$, $P$ being the polarisation of the ferromagnetic leads (assumed to be the same for the injector and the detector). Unlike in normal metals, spin accumulation in superconductors is a non-linear function of the injection voltage $V_L$, reflecting the voltage dependence of the junction conductance set by the BCS quasiparticle density of states.

The spin imbalance contribution to the nonlocal voltage can be isolated by measuring the latter when the magnetisations of injector and detector are aligned ($V_{NL}^P$) and anti-aligned ($V_{NL}^{AP}$). Inverting Equations~\ref{currents} one gets, in the limit $\mu_s\ll V_L$, $\Delta$:

\begin{eqnarray}\label{vdiff}
V_{NL}^{AP}(V_L)-V_{NL}^P(V_L) = \frac{P^2 \rho_{s}}{ g_{NS}\Omega}I_e(V_L)
\end{eqnarray}

\noindent where $I_e(V_L)$ is the local  charge current, $\rho_s = \tau_s /(N_0 e^2)$ the spin resistivity, $g_{NS}$ the normalised conductance of the detector junction and $\Omega$ the injection volume~\cite{31}. We can define a spin resistance $R_s(V_L)=[V_{NL}^{AP}(V_L)-V_{NL}^{P}(V_L)]/I_e(V_L)$. The result of Eq.~\ref{vdiff} for spin accumulation arising from a spin injection current agrees with earlier calculations by Maekawa et al.~\cite{33}, and by Zhao and Hershfeld.~\cite{32}. This signal is directly proportional to $\tau_s$. Thus, just as the electrical resistance probes the momentum relaxation time, so also the spin resistance probes  $\tau_s$.

The question arises as to the microscopic mechanisms which determine the spin imbalance relaxation time in a superconductor. In metals, spin relaxation is generally due to spin-flip scattering off magnetic impurities and/or spin-orbit scattering~\cite{36,37,38}. (Note that the spin-orbit interaction is time-reversal invariant and must therefore be combined with spatial or momentum-space inhomogeneities in order to flip spins.) In superconductors, interactions between quasiparticles and Cooper pairs also may affect spin relaxation. Calculations in the limit in which the out-of-equilibrium distribution function is practically unchanged from the equilibrium one (i.e. $\mu_s$ much smaller than all other energy scales) have shown a monotonic increase of the spin-flip time $\tau_{sf}$ for spin-orbit scattering when temperature decreases~\cite{40,41}. $\tau_{sf}$ shows instead a minimum below $T_c$ when spin-flip by magnetic impurities is the dominant relaxation mechanism~\cite{32}. (These calculations are similar to those for $T_1$ in nuclear magnetic resonance~\cite{42}.) Calculations within the quasi-classical approximation pointed to an enhanced spin-flip scattering at the gap edge leading to a strong suppression of the spin imbalance at low temperature~\cite{43}.

Early experiments on superconducting Al~\cite{44} and Nb~\cite{45} junctions of macroscopic dimensions (area $1mm^2$) showed that spin accumulation produces an excess of quasiparticles. This was interpreted as a longer recombination time due to a very long quasiparticle spin lifetime (i.e. $\tau_s\gg\tau_{in}$, the inelastic scattering time). In other words, spin relaxation was proposed as the bottleneck for quasiparticle relaxation.

\section{Spin diffusion in mesoscopic spin valves}

The long spin relaxation time measured in light metals at low temperatures (compared to that in heavy metals), and also indications of an increase of the spin lifetime in the superconducting state, suggest that the superconducting spin relaxation length could be on the order of $\mu$m, a length scale accessible to standard electron lithography. The spatial decay of spin imbalances can thus be probed directly with detector electrodes placed at different distances from the injection point. Spin-polarised electrons diffuse from the injection point (or area) and relax over a length $x_s$. When both injector and detector are ferromagnets, the device is known as a `lateral spin valve'. In superconductors, as mentioned above, charge and spin imbalances can diffuse (from the injection point) and relax over different length scales.

\begin{figure}[!h]
\centering\includegraphics[width=14cm]{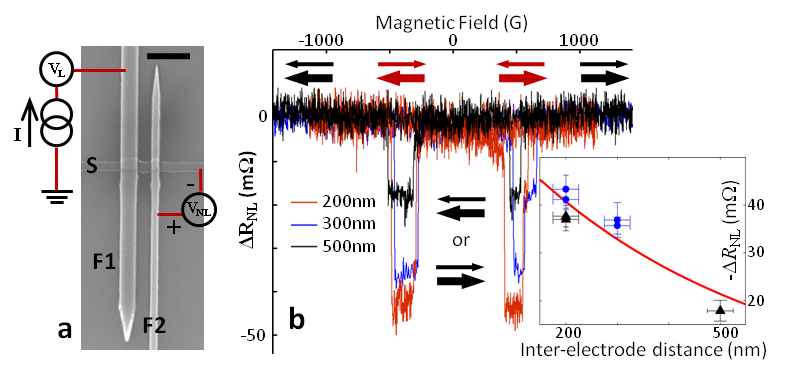}
\caption{(From Ref.~\cite{59}) \textbf{a}, Scanning electron micrograph of a spin valve device (scale bar = 1$\mu$m), and schematic drawing of the measurement setup. A current I is injected from a ferromagnet (F1, Co/Pd) into a superconductor (S, Al) through a tunnel barrier. The nonlocal voltage $V_{NL}$ and nonlocal differential resistance $R_{NL}= dV_{NL}/dI$ are measured between a distant ferromagnetic electrode (F2, Co/Pd) and S as a function of magnetic field (applied parallel to F1 and F2) and temperature; this probes the chemical potential of the spin up or down electrons with respect to the Cooper pairs depending on the relative orientations of F1, F2 and the magnetic field. The local voltage and local differential resistance $R_L= dV_L/dI$ (between S and F1) are measured simultaneously. \textbf{b}, Nonlocal magnetoresistance measurements at 4K (where the aluminium is in its normal state) allow us to identify the cohercitive fields of F1 and F2, and hence  relative alignments. (Inset) The dependence of the magnetoresistance signal on device length (distance between F1 and F2) follows the exponential decay expected from Equation~\ref{spinres}. The fit (solid line) yields to a spin flip length of 450$\pm$50 nm and a spin imbalance relaxation time $\tau_{s1}$ = 48$\pm$10 ps.}
\label{fig1}
\end{figure}

Spin imbalance in lateral spin valves depends on the distance $x$ (of the detector) from the injection point. The spin resistance decays exponentially with $x$ and is given by:

\begin{eqnarray}\label{spinres}
R_{s}(x)=  \frac{P^2 \rho_{s}}{g_{NS} A x_{s}}e^{-x/x_{s}}
\end{eqnarray}

\noindent where $A$ is the cross-sectional area of the metallic or superconducting wire perpendicular to the main direction of diffusion (The system is assumed to be quasi-one-dimensional.)

Generalising the Johnson technique to lateral spins valves Jedema et al.~\cite{47,48} showed that the spin imbalance relaxes exponentially as expected. They obtained characteristic decay lengths of 650 nm at 4.2 K and 350 nm at room temperature. This corresponds to spin imbalance relaxation times of 100 ps and 45 ps respectively in good agreement with previous experiments. A measurement of the spin resistance in a lateral spin valve and its $x$-dependence is shown in Figure~\ref{fig1}, with the Al in the normal state.

Following Johnson~\cite{49}, Jedema et al. also used the Hanle effet (coherent spin precession around a magnetic field perpendicular to the spin polarisation)~\cite{48} to directly measure the spin imbalance relaxation time.

\begin{eqnarray}\label{hanle}
R_s = \frac{P^2 \rho_{s}}{g_{NS} A x_{s}}\int_{0}^{+\infty}P(t) cos( \omega_L t) e^{-x/x_{s}}dt
\end{eqnarray}

\noindent where $\omega_L = g \mu_B H/ \hbar$, with $\mu_B$ the Bohr magneton and $H$ the applied magnetic field.
$P(t)$ is the probability that, once injected at the injector F1 (cf. Figure~\ref{fig1}), an electron arrives at the detector F2 after a diffusion time $t$. For a diffusive wire $P(t) = 1/\sqrt{4 \pi Dt} e^{-x^2/(4Dt)}$.

Unfortunately the Hanle effect cannot be measured in superconducting thin films as superfluid screening (the Meissner effect) prevents perpendicular magnetic fields from uniformly penetrating the film. In addition, (nonuniform) perpendicular magnetic fields strongly modify the density of states in the film complicating data interpretation. (Field penetration would be uniform in superconducting nanowires whose diameters are much smaller than the penetration depth, regardless of field orientation; however, we are not not aware of spin injection experiments in these systems.)

Measurements on superconducting lateral spin valves also suffer from stray fields due to the magnetisation of the ferromagnetic electrodes; this can strongly affect superconductivity in a spatially nonuniform manner. As spin resistance is defined as the difference in the nonlocal signal between the parallel and anti-parallel detector-injector magnetisation configurations, it is of the essence to be able to distinguish between out-of-equilibrium spin-dependent effects and equilibrium magnetic landscape effects.

Designed to minimize stray field effects while switching between anti-parallel and parallel configurations, the experiment performed by Poli et al.~\cite{50} yielded the nonlocal spin resistance in the superconducting state of a lateral spin-valve similar to that used by Jedema et al.~\cite{48} and Valenzuela et al.~\cite{51}. The spin relaxation length was observed to decrease in the superconducting state with respect to the normal state, consistent with spin relaxation dominated by spin-flip scattering from magnetic impurities (cf. Ref.~\cite{43}). In contrast, a recent experiment by Wakamura et al.~\cite{52} showed that the spin relaxation time increases by a factor of at least 4 with respect to the normal state in Nb, a superconductor with strong spin-orbit coupling. In this experiment, a (superconducting) wire between the injector and detector of a lateral spin valve serves as a spin current sink; this `spin absorption' technique~\cite{morota,hoffmann} allows an estimate of the spin imbalance relaxation time in materials for which $x_{s}$ is not lithographically accessible.)

\section{Spin injection by Zeeman filtering}

An external magnetic field can be used not only to control the magnetisation orientation of ferromagnetic electrodes but also to change the quasiparticle excitation spectrum in the superconductor. In thin films whose thickness $d$ is smaller than $\sqrt{(3 \hbar g \mu_B)/(D e^2 H)}$~\cite{53}, orbital screening effects due to an in-plane magnetic field $H$ can be neglected. This condition on $d$, known as the Pauli limit, comes from  imposing that the diamagnetic energy be smaller that the Zeeman energy. It results in smaller $d$ than the condition $d \ll \lambda$ the magnetic field penetration depth, which ensures a homogeneous magnetic field in the superconductor. The (dominant) Zeeman effect then shifts quasiparticle energies up or down depending on their spin orientation with respect to the applied field $E_\uparrow = E+E_Z$, $E_\downarrow = E-E_Z$, with $E_Z = \mu_B H$ the Zeeman energy~\cite{54}. The resulting spin-splitting of the BCS density of states was first observed by Tedrow and Meservey in tunnelling spectroscopy measurements~\cite{55}.

Building on the concept introduced by Huertas-Hernando et al.~\cite{56} of the `absolute spin valve', Giazotto et al.~\cite{57} pointed out that the Zeeman spitting on the BCS density of states can turn a superconductor into an excellent spin source. Current polarisations of up to 100\% have been predicted in the tunneling limit (low transparency) for a bias voltage $\Delta-E_Z < V_L <\Delta+E_Z$.

Two experiments have followed this idea~\cite{58,59}. Introducing a Zeeman-split BCS DOS in Equations~\ref{currents}; assuming $\mu_s$, $k_B T$, $\mu_B H \ll V_L$ and $x \ll x_s$; and expanding in $\mu_B H$, the spin resistance is:

\begin{eqnarray}\label{lineardev}										
R_s(V_L) = \frac{P}{2 g_{ns}\Omega}\big[P \rho_{s2}+\frac{d N_BCS(V_L)/dV_L}{N_BCS(V_L)}\mu_B H \rho_{s1} + ... \big]
\end{eqnarray}

\begin{figure}[!h]
\centering\includegraphics[width=14cm]{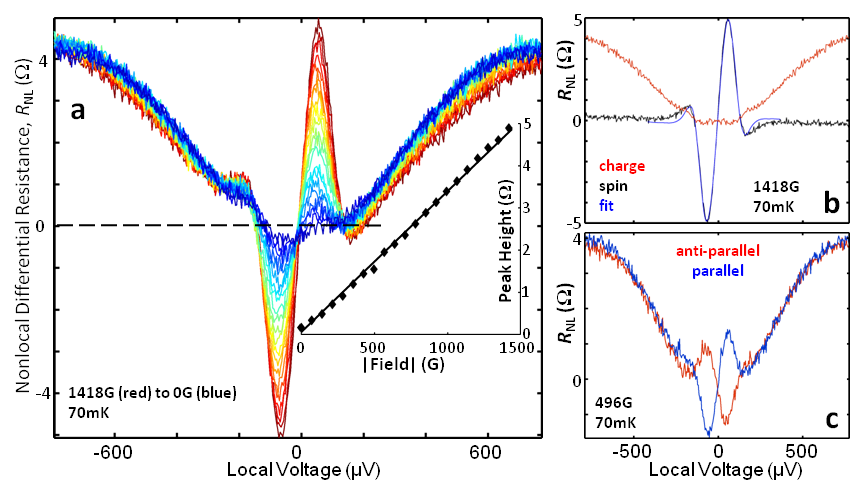}
\caption{(From Ref.~\cite{59}) \textbf{a}, Differential nonlocal resistance as a function of local voltage at different magnetic fields from 1418G (red) to 0G (blue) of the device presented in Figure~\ref{fig1} when the Al is in the superconducting state. Anti-symmetric peaks due to spin imbalance are seen on a field-independent symmetric background due to charge imbalance. (Inset) Peak height as a function of magnetic field (from anti-symmetrised data). The straight line is a guide to the eye. \textbf{b}, The anti-symmetric part of the trace at 1418G in (a), due primarily to spin imbalance. The blue line is a fit to our theory, yielding a spin imbalance relaxation time of about 25ns. The red line is the symmetric part of the trace at 1418G (a), due primarily to charge imbalance. \textbf{c}, Differential nonlocal resistance as a function of local voltage at 496G with the detector electrode aligned (blue line) then anti-aligned (red line) with the injector electrode and the magnetic field. The spin imbalance signal changes sign while the charge imbalance signal remains the same. The difference in amplitudes between the two spin signals is due to a residual magnetic field. (See Supp. Info. of Ref.~\cite{59}).}
\label{fig3}
\end{figure}

This expression is particularly suggestive: the spin imbalance can be seen here to depend clearly on the polarisation of the injector electrode (first term) as well as on the Zeeman splitting of the DOS in the superconductor (second term). Here we have introduced two different spin resistivities $\rho_{s2}$ and $\rho_{s1}$ associated with two different relaxation times $\tau_{s2}$ and $\tau_{s1}$ in the first and second terms, respectively. Physically this corresponds to different relaxation mechanisms for spin imbalances in the presence or absence of a Zeeman-induced splitting of the BCS DOS. The shape of the signal measured in the experiment (Figure~\ref{fig3}.) and its amplitude as a function of the applied magnetic field in the limit of small field agrees well with Eq.~\ref{lineardev} provided the off-set produced by the first term is negligible. $\tau_{s1}$ obtained from the data is 10ns  much longer than the spin relaxation time measured in the normal state. Furthermore $\tau_{s1} \gg \tau_Q$~\cite{60} consistent with spin-charge separation as suggested theory~\cite{32}.

\begin{figure}[!h]
\centering\includegraphics[width=14cm]{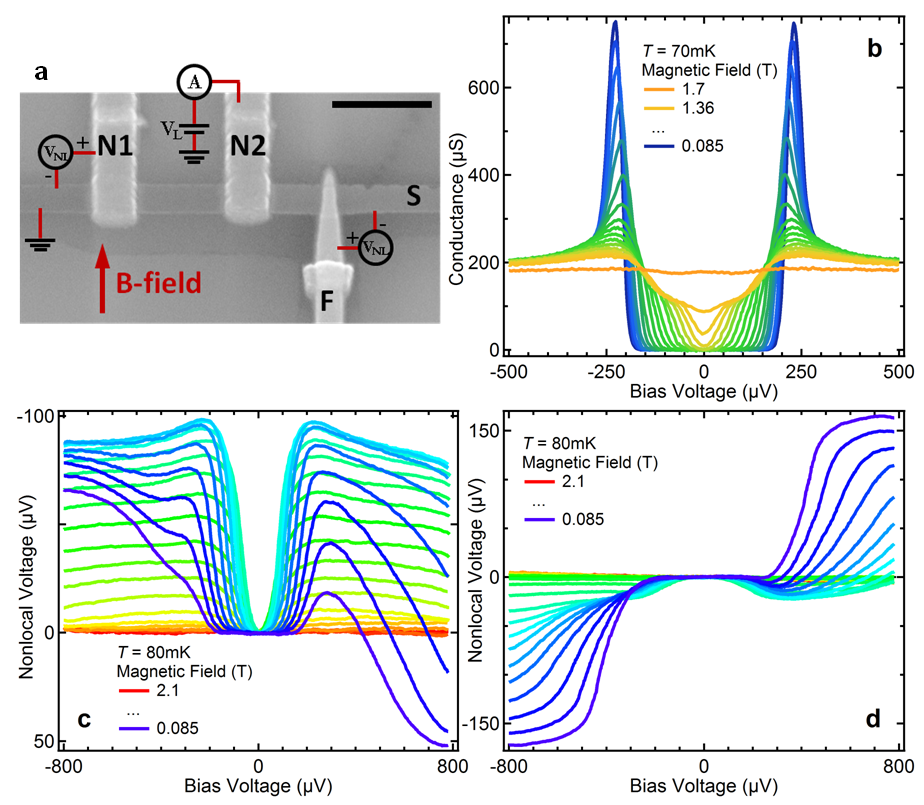}
\caption{(From Ref.~\cite{65} except for (b).)  \textbf{a}, Scanning electron micrograph of a device similar to that shown in Figure~\ref{fig1}a, with a normal metal injector (N2) and two non-local detection junctions (N1, F). (Scale bar = 1$\mu$m.) \textbf{b}, In a preliminary measurement, the differential conductance $G_L=dI/dV_L$ across the junction N1-S, as a function of the bias voltage $V_L$, at different values of the applied magnetic field $H$. The Zeeman splitting is visible in $G_L$, which is proportional to the density of states in S. The next two panels show the non-local voltage $V_{NL}$ measured at a ferromagnetic detector (\textbf{c}) and at a normal metal detector (\textbf{d}) as a function of the (local) bias voltage $V_L$ across N2-S for different $H$. Here the injector (N2) is Al (100nm) in the normal state. The amplitude of the signal from spin imbalance (even signal) initially grows with increasing field, saturates and then gradually vanishes at the critical field. The amplitude of the charge imbalance (odd signal) instead decreases monotonically with increasing field.}
\label{fig2}
\end{figure}

Note that, as spin polarisation is provided mainly by the Zeeman splitting of the superconducting density of states, the injector electrode no longer needs to be a ferromagnet. This can be seen in Figure~\ref{fig2}, where we present data from a device, similar to that shown in Figure~\ref{fig1}, but in which the injector is normal. The thickness of the superconducting Al wire is reduced to 8.5 nm to increase the critical field from $\sim$650 mT to 2100 mT. While the spin accumulation first increases with increasing field (Figure~\ref{fig2}c), the charge imbalance monotonically vanishes with field as expected for increasing orbital depairing (Figure~\ref{fig2} d). Note also that the signals from spin and charge imbalance have different symmetries, even and odd respectively, in agreement with Equation~\ref{currents}.

Spin currents carried by quasiparticles can be converted into transverse charge currents by the inverse spin Hall effect~\cite{61}, resulting in charge accumulation at the transverse edges of the superconductor~\cite{62}. The spin Hall effect originates from side jump and skew scattering induced by spin-orbit coupling~\cite{zutic,hoffmann,sinova}. Quasiparticle-mediated spin Hall and inverse spin Hall effects are expected to be greatly enhanced in the superconducting state compared to the normal state~\cite{62}, as the nonlocal spin resistance is directly proportional to the quasiparticle density that decreases exponentially with temperature below $T_c$, as $e^{-\Delta /k_B T}$. Very recently an increase of three orders of magnitude in the nonlocal inverse spin Hall resistance of a NbN wire well below $T_c$ (compared to the normal state) was reported~\cite{63}. The inverse Hall signal was observed only when injector and detector electrodes were closer together than the charge relaxation length.

\section{Spin-dependent even and odd modes}

The experiments by Quay et al.\cite{59} and Hübler et al.~\cite{58} show unequivocally that a magnetization appears in a mesoscopic superconductor in the Pauli limit when the superconductor is driven out-of-equilibrium by a tunneling current from both a ferromagnetic and a normal electrode (see also Ref.s~\cite{64} and~\cite{65}).

These experiments are consistent with --- and were interpreted in the framework of --- the spin imbalance model presented above, in which spin imbalance appears as a chemical potential shift $\mu_s$ which is opposite for opposite spins, thus bracketing out the question of the spin-dependent energy distribution function.

\begin{figure}[!h]
\centering\includegraphics[width=13cm]{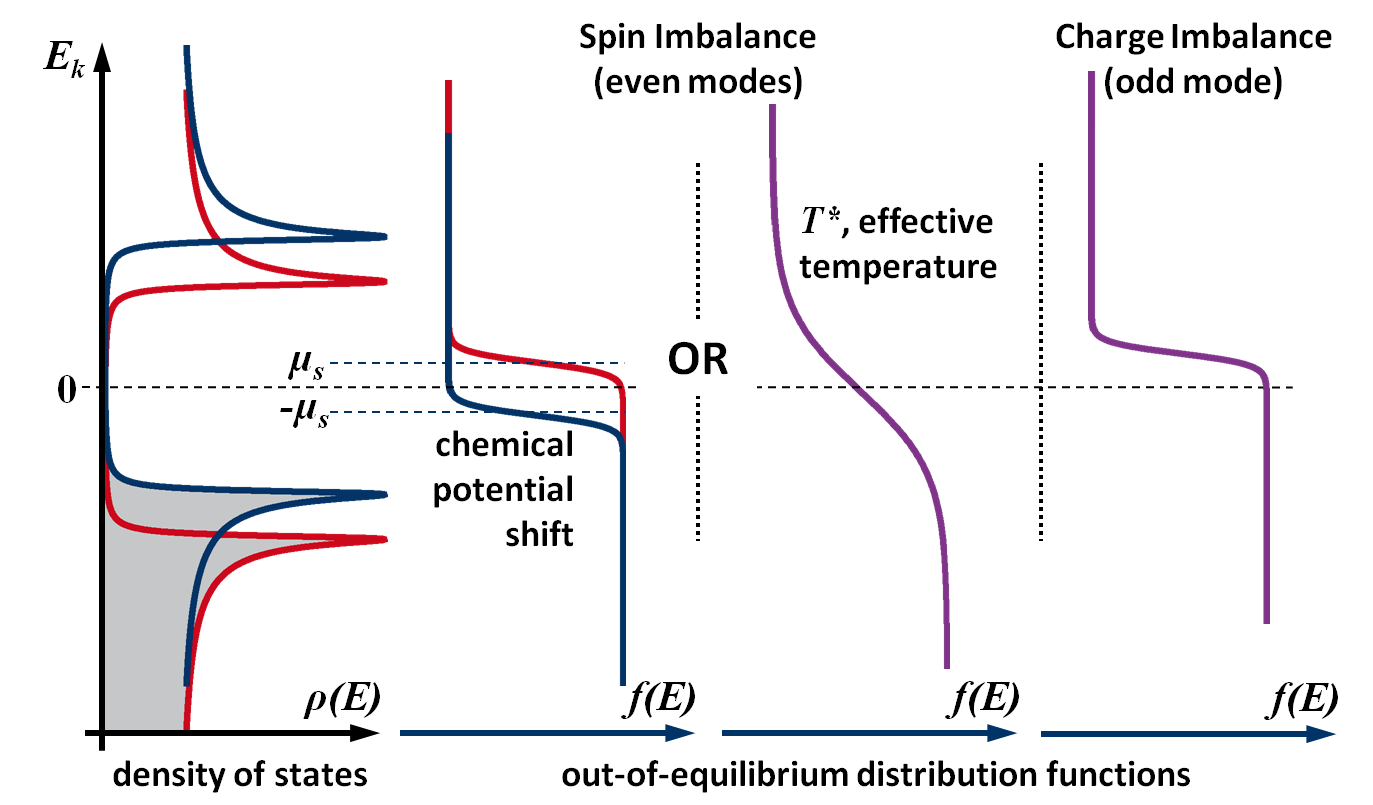}
\caption{Spin and Charge imbalances. (All parnels except rightmost) In a simple picture, two sources of spin imbalance in a Zeeman-split superconductor: a (symmetric) spin-dependent shift in the chemical potential, or a spin-independent effective temperature different from that of the rest of the system (phonons, condensate and electrons in other electrodes). A more general description would be based on spin-dependent chemical potentials and distribution functions. (Leftmost and rightmost panels) A charge imbalance occurs when there are unequal numbers of hole- and electron-like excitations, e.g. when the spin-independent chemical potential is different from that of the pairs.}
\label{fig4}
\end{figure}

In a simple picture, non-equilibrium excitations in Zeeman-split superconductors can be described by four parameters together quantifying the charge and spin degrees of freedom of the system: $\mu_\downarrow$, $\mu_\uparrow$, $T_\downarrow^*$ and $T_\uparrow^*$. This presupposes that the electron-electron interaction time $\tau_{e-e}\ll\tau_{e-ph}$ the electron-phonon time, and that the distribution functions can thus be approximated by Fermi-Dirac distributions, and generalises the even and odd modes of the non-Zeeman-split case by taking spin into account. The conceptually simplest spin imbalances are then due to (or associated with) either $\mu_\downarrow =\mu_s$ and $ \mu_\uparrow = -\mu_s$ as in Ref.s~\cite{58,59,34}, or $T_\downarrow^* = T_\uparrow^*$ and both different from the equilibrium (bath) temperature (Figure~\ref{fig4}). In both cases, a finite voltage will be measured between the superconductor and a ferromagnetic detector with which it is in contact via a tunnel barrier. The latter scenario, which corresponds to a spin-dependent thermoelectric effect, is a result of the broken electron-hole symmetry of the DOS of each spin in the superconductor due to the Zeeman field.

A complete description of the non-equilibrium excitations requires the knowledge of the spinful distribution functions $\tilde{f}_\downarrow(E)$ and $\tilde{f}_\uparrow (E)$, where the $\tilde{f}(E)$ need not be Fermi-Dirac not share the same chemical potential.



Indeed, by solving Boltzmann diffusion equations, and taking into account local modification of superconductivity due to the out-of-equilibrium distribution functions, Krishtop et al.~\cite{66} found that nonlocal signals with similar bias dependence to those measured in Ref.s~\cite{58,59} are also expected in the second case above~\cite{66}, as well as the more general case where $f_\downarrow(E) = f_\uparrow(E)$ is not a Fermi-Dirac distribution. Bobkova et al.~\cite{68} and Silaev et al.~\cite{67} find similar behaviour for the nonlocal signal using the quasiclassical Keldysh-Usadel formalism. (See also Ref.s~\cite{machon,ozaeta}.)

Silaev et al.~\cite{67} further proposed a description of the out-of-equilibrium state in the following basis: charge, spin, energy and spin energy, with spin and energy modes contributing to the observed signal. The former relaxes via elastic spin-flip processes; the latter relaxes via inelastic processes (electron-electron or electron-phonon scattering) and is unaffected by elastic scattering.

All of this recent theoretical work point to the electron-phonon relaxation time as an upper bound for the spin imbalance relaxation time. This is a natural explanation for the long times measured by Quay et al.\cite{59} and Hübler et al.~\cite{58}: however, the precise `modes' excited and out-of-equilibrium distribution functions produced remain to be experimentally investigated.

The spin imbalance relaxation length is observed to increase with the applied magnetic field, particularly when the orbital depairing is strong, and is explained by Bobkova et al.~\cite{68} as follows: Quasiparticle scattering occurs primarily at the gap edge with rates which go as $E^2$ for electron-electron scattering and $E^3$ for electron-phonon scattering. A reduced gap results in a diminished inelastic scattering rate and thus an increased lifetime. The upper limit for the spin relaxation time becomes the normal state energy relaxation time at that temperature. However this characteristic time scale corresponds to the lifetime of the out-of-equilibrium magnetization in the superconductor rather than the spin-flip time of each quasiparticle. This is similar to $T_1$ in nuclear magnetic resonance terminology~\cite{42}.

\begin{figure}[!h]
\centering\includegraphics[width=14cm]{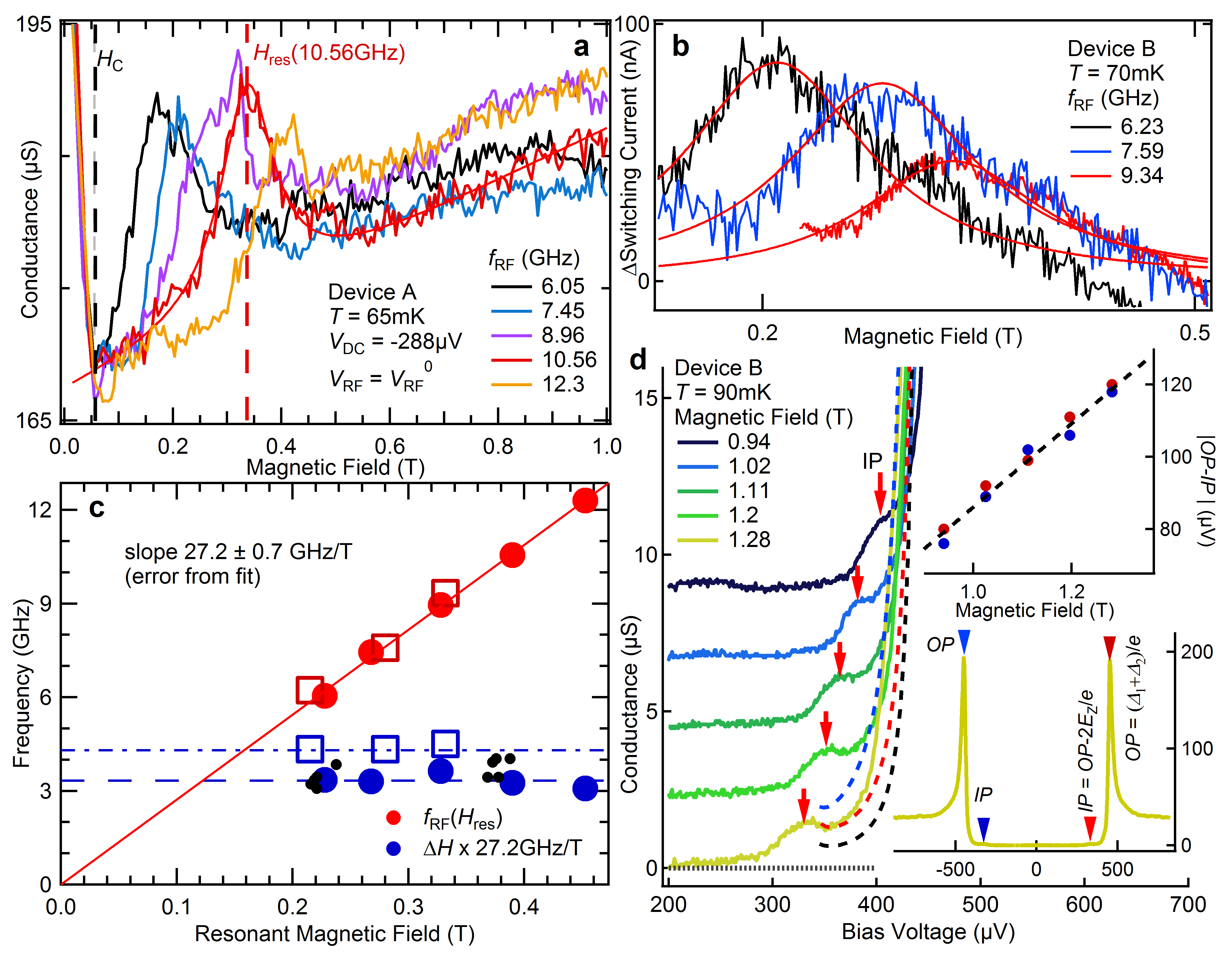}
\caption{(From Ref.~\cite{72}) Data from quasiparticle spin resonance experiments. Device A is identical to that shown in Figure~\ref{fig2}a: A static magnetic field, $H$ is applied parallel to the superconducting bar (S, Al) and a sinusoidal signal of rms amplitude $V_{RF}$ and frequency $f_{RF}$ in the microwave range applied across the length of S (with a lossy coaxial cable in series), resulting in a high-frequency field perpendicular to $H$. S is 8.5nm (6nm) thick in Device A (B). In the first detection scheme, a voltage $V_D$ is applied between S and a normal electrode (N1, thick Al) with which it is in contact via an insulating tunnel barrier (I, Al$_2$O$_3$). The differential conductance $G_D=dI/dV_D$ is measured, where $I$ is the current between N1 and S. In the second detection scheme, the switching current $I_s$ of S is measured. \textbf{a}, NIS junction conductance $G$ as a function of $H$ at $V_D$ = -288$\mu$V and constant $V_{RF}$ for different $f_{RF}$. The black vertical line indicates the critical field of N. $H_{res}$ and $\Delta H$ are obtained for each $f_{RF}$ by fitting a Lorentzian with a linear background. The fit for $f_{RF}$ = 10.56GHz is shown (thin red line) and $H_{res}$ indicated with a red vertical line. $I_s$ measurements (cf. Ref.~\cite{72}) on the same device give the same result. \textbf{b}, $I_s(H)$ of S for different $f_{RF}$. \textbf{c},  $H_{res}$ and $\Delta H$ the resonance linewidth (full-width at half-maximum) as a function of $f_{RF}$ (red and blue markers respectively). A linear fit to the red circles gives a Landé g-factor of 1.95$\pm$0.2. The black dots indicate values obtained at different powers or with the second detection scheme. (See Ref.~\cite{72} and Supp. Info.) The circles (squares) are values obtained from Device A (B). \textbf{d}, $G(V)$ across an SIS' junction, at different magnetic fields $H$ (offset by 2.2$\mu$S). Apart from the principal, outer peak at $V=(\Delta+\Delta')/e$, with $\Delta$ ($\Delta'$) the superconducting energy gap of S (S'), a smaller, inner peak can be seen at $\sim V=(\Delta+\Delta'-2E_Z)/e$. Fitting the data at $H$=1.28T to numerical calculations based on Ref.s~\cite{53,54} (red dotted line) yields a spin-orbit time $\tau_{so}$ of 45$\pm$5ps. Numerical results for $\tau_{so}$ = 23ps and 69ps are also shown (blue and black dotted lines respectively). Lower Inset: Full conductance trace at $H$ = 1.28T, showing all peaks. Upper Inset: Distance in $V$ between outer and inner peaks at positive (red dots) and negative (blue dots) energies. The black dotted line, which has a slope of $2E_Z/e$, is a guide to the eye.}
\label{fig5}
\end{figure}

\section{Spin resonance}

Spin resonance experiments allow a direct measurement of the spin coherence time, $T_2$~\cite{69,71}. In a typical electron spin resonance (ESR) experiment, electrons are immersed in an external homogenous static magnetic field, $H$. Microwave radiation creates a perturbative transverse magnetic field (perpendicular to the static field) of frequency $f_{RF}$. The power $P(H, f_{RF})$ absorbed by the spins from the microwave field is determined, usually by measuring the fraction of the incident microwaves that is not absorbed, i.e. either transmitted or reflected~\cite{70}. When $H$ is tuned to its resonance value, $H_{res} = 2 \pi f_{RF} / \gamma$ \textemdash with $ \gamma $ the gyromagnetic ratio \textemdash  the electron spins precess around $H$ and $P(H,f_{RF})$ is maximal. $T_2$ if proportional to the full-width at half-maximum of the resonance~\cite{69}.

In a very recent experiment Quay et al.~\cite{72} measured the spin resonance of out-of-equilibrium quasiparticles using two `on-chip', local microwave powermeters: a supercurrent measurement and the conductance across a normal metal-insulator-superconductor junction. At resonance, microwave radiation is absorbed by the precessing out-of-equilibrium quasiparticle magnetization.

Some data from Ref.~\cite{72} are shown in Figure~\ref{fig5}. Figures~\ref{fig5}a and Figure~\ref{fig5}b show spin resonances obtained for two devices with superconducting Al bars of different thicknesses ($d_A = 8.5nm$ and $d_B = 6nm$). The resonant field as a function of the microwave frequency from these and other measurements is reported in Figure~\ref{fig5}c. A linear fit to the data gives a g-factor of 1.95$\pm$0.2, consistent with previous measurements of electrons in Al in the normal state (Figure~\ref{fig5}c). The spin relaxation time $T_2$ is also shown in Figure~\ref{fig5}c. $T_2$ = 95$\pm$20ps (70$\pm$15ps) for Al which was 8.5nm (6nm) thick. Both the order of magnitude of $T_2$ , as well as the fact that it is inversely proportional to the film thickness, are consistent with spin coherence limited primarily by the Elliott-Yafet spin-orbit scattering time $\tau_{so}$~\cite{36,37}. Further, these figures agree with an independent, equilibrium measurement of $\tau_{so}$ (Figure~\ref{fig5}d).

We note that, unlike normal metals where $T_2=T_1$ the spin imbalance relaxation time, in superconductors $T_1$ is much larger than $T_2$, and that $T_2$ is of the order of the spin imbalance relaxation time measured in the normal state~\cite{48,59}, consistent with plural sources and relaxation mechanisms of spin imbalance in the superconducting state~\cite{73}.

\section{Influence of spin imbalance on the superconducting condensate}

The influence of out-of-equilibrium quasiparticle distributions, including spin imbalance, is in principle accounted for by the self-consistent BCS gap equation~\cite{39}:

\begin{eqnarray}\label{gapeq}
\frac{1}{N_0 V_{BCS}} = \int_\Delta^{\hbar\omega_D} dE \big[1-f_\uparrow(E)-f_\downarrow(E)\big]/E
\end{eqnarray}

\noindent where $\omega_D$ is the Debye energy and $V_{BCS}$ the BCS interaction strength.

The solution in the presence of a Zeeman/Pauli (as opposed to orbital) magnetic field was given by Sarma in 1963~\cite{19}. In the case where a spin imbalance can be described by $\mu_s$, while the quasiparticle temperature is the bath temperature, $\mu_s$ acts as an effective magnetic field and simply adds to the applied field $\mu_B H$~\cite{75,bobkova-prb,jin}. The effects of the two thus add or subtract depending on their relative sign. The case of cancellation has been described as a `recovery' of superconductivity through spin injection~\cite{bobkova-prb}. We note, however, that in the case described here, where spin imbalance is provided by Zeeman-splitting due to the same magnetic field, the two effects have the same sign and there is no recovery.

More generally, as already pointed out above, the non-equilibrium spin-dependent distribution functions due to spin injection have to be calculated by solving kinetic equations, which take spin diffusion into account. This is particularly important for mesoscopic devices, as quasiparticles thermalise internally, which is to say recover a Fermi-Dirac distribution, over distances larger than the electron-electron scattering length (typically of the the order $\mu$m at very low temperature)~\cite{6}.

In order to observe the predicted modification of the gap due to spin imbalance, a number of experiments have been performed on high-Tc superconductors. Indeed, it has been predicted that spin injection results in a greater suppression of d-wave superconductivity than of conventional s-wave superconductivity, mainly due to injection of low-energy excitations at the order-parameter nodes~\cite{80}. Most experiments in this direction have focused on reductions of the critical current with spin injection~\cite{74}. V.A. Vas'ko et al.~\cite{77} showed complete suppression of superconductivity in La$_{2/3}$Sr$_{1/3}$MnO$_3$/La$_2$CuO$_4$/DyBa$_2$Cu$_3$O$_7$ heterostructures. Similar results have been obtained in Au/YBa$_2$Cu$_3$O$_7$/LaAlO$_3$/Nd${_0.7}$Sr$_{0.3}$MnO$_3$ heterostructures~\cite{78,79}. There is nevertheless some debate around these results, and in particular whether they are due to $f_\uparrow(E) \neq f_\downarrow(E)$ or simply to spin-independent out-of-equilibrium quasiparticles~\cite{gim}. To our knowledge, there has not yet been a spin-resolved measurement of the distribution function correlated with spectroscopic evidence of gap reduction by spin injection.

Coming back to the case of fast thermalization (i.e. quasiparticles thermalise quickly amongst themselves and with the environment compared to the elastic spin-flip time, and the spin imbalance is given by $\mu_s$), at zero temperature the gap is independent of $\mu_s$ for $\mu_s<\Delta_0$ while at finite temperature the system goes through a first order transition to the normal state at $\mu_s < \Delta_0$~\cite{75,81}. In addition, Sarma pointed out that for Zeeman-split superconductors at zero temperature the self-consistent gap equation admits a solution corresponding to conventional Cooper pairs occupying the energy region $|E|<E_F-\sqrt{H^2-\Delta^2}$ while a shell $|E|<\sqrt{H^2-\Delta^2}$ around the Fermi surface is occupied only by spin down electrons~\cite{19}. This solution appears as a mixture of Pauli paramagnetism and a BCS superconductor with a field-dependent gap.

The Sarma state and the related `breached pair'~\cite{liu} state have received a great deal of attention in the context of BCS superfuidity in cold atoms where the imbalance between spin up and down densities can be finely controlled~\cite{82,83,84}. Evidence of a core-shell configuration where the BCS superfluid occupies the core while the shell is a gas of spin-polarised atoms has been shown by microscopy~\cite{85}. There has, however, yet to be theoretical or experimental work on inhomogeneous superconducting states in which spins and Cooper pairs are separated in real or momentum space. We note that, in addition to cold atoms, BCS-like theories with spin/population imbalance have also been applied to dense quark matter~\cite{casalbuoni,shovkovy}; they are thus of quite general interest.

\section{Conclusions}

Out-of-equilibrium superconductivity has been an active area of research since the 1970s; however, comparatively little attention has been paid to spinful excitations. The past few years have seen a revival of interest in the latter, in the wake of experiments demonstrating out-of-equilibrium magnetisation, spin-charge separation and quasiparticle resonance in superconducting aluminium in the Pauli limit. With close analogues in cold atom systems and dense quark matter, spin imbalanced superconductors are of interest to the broader physics community beyond condensed matter.

Nevertheless, comprehensive theoretical study of out-of-equilibrium superconductivity where spin degeneracy is lifted by Zeeman splitting, which includes the self-consistent calculation of the spatial dependence of the superconducting gap, is still a work in progress. Apart from traditional quasiclassical approaches, attempts to apply hydrodynamical methods to the problem are also underway~\cite{shpielberg}. Among open questions are the mutual interactions between the condensate and spinful excitations and the existence of novel ground states, including those due to the influence of spin imbalances on the internal spin structure of the superconducting order parameter~\cite{zapata,ho}.

\aucontribute{Both authors contributed equally to this manuscript; all data shown have been published elsewhere (see citations).}

\competing{The authors declare that they have no competing interests.}

\funding{The authors' and collaborators' work mentioned in this review was funded by European Research Council Starting Independent Researcher (NANO-GRAPHENE 256965) and Synergy Grants, a C'NANO grant (DYNAH) from the Ile-de-France region, ANR Blanc grants (DYCOSMA and MASH) from the French Agence Nationale de Recherche, and the Netherlands Organization for Scientific Research (NWO/OCW).}

\ack{The authors' work mentioned in this review benefitted from collaboration with Cristina Bena, Denis Chevallier, Yann Chiffaudel, Clément Dutreix, Marine Guigou, Christoph Strunk, Mircea Trif and Maximilian Weideneder. We also acknowledge helpful discussions with Detlef Beckmann, Jaroslav Fabian, Julien Gabelli, Manuel Houzet, Hervé Hurdequint, Tatiana Krishtop, Brigitte Leridon, Jérôme Lesueur, Julia S. Meyer, Yuli Nazarov, Bertrand Reulet and Stanislas Rohart.}


\end{document}